\documentclass[fleqn,usenatbib]{mnras}

\usepackage{newtxtext,newtxmath}

\usepackage[T1]{fontenc}

\DeclareRobustCommand{\VAN}[3]{#2}
\let\VANthebibliography\thebibliography
\def\thebibliography{\DeclareRobustCommand{\VAN}[3]{##3}\VANthebibliography}

\usepackage{graphicx}	
\usepackage{amsmath}	
\usepackage{parskip}
\usepackage{enumerate}

\title{Detection of new pulsars at the frequency 111 MHz}

\author[S. A. Tyul'bashev et al.]{
S.A. Tyul'bashev,$^{1}$\thanks{E-mail: serg@prao.ru (SAT)}
V.S. Tyul'bashev,$^{2}$
V.V. Oreshko,$^{1}$
S.V. Logvinenko$^{1}$
\\
$^{1}$ P.N. Lebedev Physical Institute of the Russian Academy of Sciences, Astro Space Center, Pushchino Radio Astronomy Observatory,\\
Radiotelescopnaya 1a, Moscow reg., Pushchino, 142290, Russia \\
$^{2}$ Faculty of Computational Mathematics and Cybernetics,
Lomonosov Moscow State University, Moscow, Russia\\
}

\begin{document}
\label{firstpage}
\pagerange{\pageref{firstpage}--\pageref{lastpage}}
\maketitle

\begin{abstract}
The pulsar search was started at the radio telescope LPA LPI at the frequency 111~MHz. The first results deals of a search for right ascension $0^h-24^h$ and declinations $+21^{\circ} - +42^{\circ}$ are presented in paper. The data with sampling 100 ms and with 6 frequency channals was used. It were found 34 pulsars. Seventeen of them previously been observed at radio telescope LPA LPI, and ten known pulsars has not previously been observed. It were found 7 new pulsars.\\
\end{abstract}




\section{Introduction}

Since the discovery of pulsars in 1967 (\citeauthor{Hewish1968}, \citeyear{Hewish1968}), searches for new pulsars have been regularly conducted in both the northern and southern hemispheres. As a rule, the search for pulsars is carried out under "lanterns": the plane of the Galaxy, globular clusters, remnants of supernova bursts. Only very few surveys cover areas exceeding at least one steradian. Of the first surveys covering a significant part of the sky, we can note the surveys of Molonglo (\citeauthor{Large1971} (\citeyear{Large1971}), \citeauthor{Manchester1978} (\citeyear{Manchester1978})), Jodrell Bank  (\citeauthor{Davies1972} (\citeyear{Davies1972}), \citeauthor{Davies1973} (\citeyear{Davies1973})), Green Bank (\citeauthor{Damashek1978} \citeyear{Damashek1978}). Of the latest, still ongoing surveys, we note AO327 on the 300-meter radio telescope in Arecibo  (\citeauthor{Deneva2013} \citeyear{Deneva2013}), GBNCC on the 100-meter radio telescope in Green Bank  (\citeauthor{Boyles2013} \citeyear{Boyles2013}), NHTRU on the 100-meter telescope in Effelsberg (\citeauthor{Barr2013} \citeyear{Barr2013}), HTRU on the 64-meter telescope in Parks (\citeauthor{Keith2010} \citeyear{Keith2010}), LOTAAS on the LOFAR  aperture synthesis system (\citeauthor{Coenen2013} \citeyear{Coenen2013}). In these surveys, years of observations are required to cover most of the celestial hemisphere. The reduction of the total observation time is achieved, as a rule, due to a small integral accumulation time in a given direction. Multiple observations of selected sites are not practiced. At the same time, it is known that pulsars are objects with strong variability caused by both external (interstellar scintillating, see, for example, review (\citeauthor{Rickett1977} \citeyear{Rickett1977}) and internal causes (flare pulsars: example pulsar J0946+0951  (\citeauthor{Vitkevich1969} \citeyear{Vitkevich1969})). Therefore, with regular observations of the entire celestial sphere, relatively strong new pulsars can be detected in previously studied areas.

On the Large Phased Array (LPA) of Lebedev Physics Institute (LPI) radio telescope, a daily sky survey is conducted in test mode, covering a full day in right ascension and $50 ^{\circ}$ in declination. Brief reports on the discovery of 7 new pulsars in this survey are published in  \citeauthor{Tyulbashev2015a} (\citeyear{Tyulbashev2015a}), \citeauthor{Tyulbashev2015b} (\citeyear{Tyulbashev2015b}). In this paper, the details of the search for pulsars are given.

\section{Observations and processing of observations}

LPA LPI which is a phased array, was upgraded in 2012. In the course of its modernization, the low-noise amplifiers of the first floor were replaced, the Butler phasing matrices were replaced, and new cable systems were laid. As a result of this work, two independent diagram-forming systems appeared on the radio telescope (LPA1 and LPA2). The first phasing system is based on the old diagram formation system. It forms 16 beams that can be shifted by half the width of the radiation pattern so that in two days of observations it becomes possible to overlap the sky with the beams of the LPA LPI at a power level of 0.8. This system of 16 beams overlaps in declination about $8^{\circ}$, and can switch so that observations are provided at declinations from $-27^{\circ}$ to $+88^{\circ}$. On LPA1, observations of pulsars are carried out, among other things, on specialized "pulsar" receivers. The maximum effective area of this system is $20,000 \pm 2000$ sq.m.. The second antenna pattern is constructed in a special way. 128 non-switchable beams have been created in it. The overlap of the beams is made according to the power level of 0.4. This beam system allows you to observe sources with declinations from $-8^{\circ}$ to $+55^{\circ}$. The effective area of LPA2, reduced to the zenith, is $47,000 \pm 2500 $ sq.m. Both LPA1 and LPA2 have a full frequency band of 2.5~MHz. A multi-channel digital receiver has been made for the LPA2, allowing for the registration of the signal from 96 beams of the LPA. These beams overlap declinations from $-8^{\circ}$ to $+42^{\circ}$.

A special system has been developed to calibrate the power of incoming signals. It provides measurement of the main parameters of the radio telescope: the noise temperature of the system, the effective area of the antenna, checking the operability of the distributed amplification system and its individual elements. The input of low-noise amplifiers switches between the antenna and the calibration noise generator. The noise generator generates 2 levels of the calibration signal corresponding to the noise temperature of the matched load and the noise temperature of the switched-on generator. The noise temperature of the matched load is equal to the ambient temperature. The temperature of the noise signal of the calibration generator is 2400 K and little depends on the ambient temperature. The measured changes in the step height do not exceed $\pm 3\%$ when the ambient temperature changes from $-15^{\circ}$ to $+43^{\circ}$ Celsius.

The multichannel digital radiometer consists of two recorders - industrial computers with multichannel receiving and recording modules that provide signal registration for 96 beams of the radio telescope. Each recorder includes six 8-channel digital signal processing modules. 4 paired ADCs~TI~ADS62P29 are installed at the input of the module. Digital processing is performed using programmable logic device PLD EP3SL780C3 (Stratix III, Altera). Used: the method of direct digitization of the signal, digital filtering systems, frequency transfer, spectral analysis. The digitization frequency is 230 MHz, the band of the recorded signal in each channel is 2.5 MHz with a central frequency of 110.25 MHz. PLD resources make it possible to implement 8 independent video converters on one chip, filtering of high-frequency and low-frequency signals, spectral analysis and processing of 8 independent data streams. The recorder module has the ability to record signal power to a hard disk in 32 or 6 spectral channels with a frequency resolution of 78 or 415 kHz, respectively. The time resolution of the signal is 12.5 ms or 100 ms.

Since July 2014, simultaneous data recording has been carried out, both with a sampling 100~ms in 6 frequency channels ("short data") and with a sampling 12.5~ms in 32 frequency channels ("long data") at hourly intervals. The time service is monitored at the beginning of each observation hour. The start of digitization of the first point is carried out with an accuracy of at least 5 ms. The accuracy of channel polling within an hour is determined by the accuracy of the quartz oscillator of the digital receiver. The estimated maximum possible time discrepancy in the hourly interval is $\pm 25$ ms ($\pm$ two points of primary "long data") and $\pm 100$~ms ($\pm$ one point of primary "short data").

To search for pulsars, a program has been written in Qt/C++ distributed under the GPL V3.0 (\citeauthor{TyulbashevV2015} \citeyear{TyulbashevV2015}), consisting of two blocks. The first block performs a direct addition of periods with a search of dispersion measures (DM). All detected periodic signals related to signal to noise $S/N \ge 4$ are recorded in the base directory. In the second block, the detected signals are analyzed. The program allows you to compare catalogs for all processed days by a number of parameters and perform additional digital filtering. For example, you can track the repeatability of periodic signals from day to day. Check the approximate coincidence of the maxima of accumulated pulses when searching for pulsars with a double period. Identify candidates for pulsars with S/N greater than the specified one. Remove known pulsars from directories. Construct dynamic spectra. There are other digital filters in the program.

The full passage of the source through the meridian of LPA LPI takes $425s/cos(\delta)$ ($\delta$ - declination of the source). The maximum sensitivity in a given direction is achieved at the moment when the meridian is crossed by the source. When processing observations, the region $\pm 1.5-2$ min from the moment of crossing the meridian is usually used. Therefore, in order to achieve maximum sensitivity, three-minute intervals with a shift of 1.5 minutes were processed during the search. Such a shift makes it possible to ensure that in the processed recordings, the pulsar will necessarily pass through the top of the antenna radiation pattern, where maximum sensitivity will be achieved.

\section{Pulsar detection}

The sensitivity when detecting extremely weak pulsars, according to the flux density is determined by the well-known equation:

$$S_{min} = \frac{S/N_{min} T_{sys}}{G \sqrt{n_p \Delta t \Delta \nu_{MHz}}} \sqrt{\frac{W_e}{P-W_e}}\, (mJy),$$

where the expected sensitivity ($S_{min}$) when searching for pulsars on full-turn antennas is determined by a given $S/N_{min}$ (in practice $S/N_{min} =6-8$), the temperature of the system ($T_{sys} = T_b + T_r$, where $T_b$ is the background temperature, and $T_r$ is the receiver temperature), the parameter $G$ associated with the normalization coefficient with the effective area of the antenna, the total accumulation time ($\Delta t$), the frequency band ($\Delta\nu$), the number of observed linear polarizations ($n_p$), the ratio of the pulsar pulse duration ($W_e$) to the pulsar period ($P$).

In the case of LPA which is an antenna array, it is necessary to take into account a number of additional factors affecting the sensitivity: a) the sensitivity of the antenna decreases with the zenith distance; b) the total observation time is limited by the time the source passes through the meridian. The maximum sensitivity at the top of the radiation pattern of the LPA LPI for 3 minutes; c) the position of the beams is fixed in declination, and therefore, if the coordinates of the source do not coincide with the coordinate of the center of the beam, the sensitivity decreases.

Let's evaluate the sensitivity of the LPA LPI when searching for pulsars. It depends, first of all, on the temperature of the galactic background and on how close to the center of the antenna array pattern the pulsar will pass, crossing the celestial meridian. Assuming a temperature in the plane of the Galaxy of 1500K, and outside the plane of 500K (see radioisophotes at 178 MHz  (\citeauthor{Tartle1962} \citeyear{Tartle1962}), which estimated the temperature and recalculated at 111 MHz with a spectral index $\alpha = -2.55$, where $S\sim \nu^{-\alpha}$), the pulsar hitting exactly in the center of the diagram (beam) and in the middle between the two beams, it is possible to estimate the best ($S_{best}$) and worst ($S_{worst}$) expected sensitivity according to the formula given above for the case when the sampling is less than the pulse duration of the pulsar. The following parameters were used for the estimates: $S/N=6$, $G=17$ K/Jy, $n_p=1$, $\Delta\nu =2.5$~MHz, $\Delta t = 180$~s, $W_e=0.1P$, $T_r= 300$K. Table~\ref{tab:tab1} gives estimates of the sensitivity of LPA for the best, worst and expected typical case.

\begin{table}
	\centering
	\caption{The expected sensitivity when searching for pulsars on LPA LPI in the direction of the zenith.}
	\label{tab:tab1}
	\begin{tabular}{cp{1cm}p{1cm}p{1cm}}
		\hline
Sensitivity & $S_{best}$ (mJy)  & $S_{worst}$ (mJy) & $S_{typical}$ (mJy)\\
		\hline
The plane of the Galaxy           &  9.9   &  24.5   &  15-20  \\
\hline
Outside the plane of the Galaxy   &  4.4   &  10.8   &  6-8  \\ 
		\hline
	\end{tabular}
\end{table}

The currently conducted pulsar search surveys (see the links in the Introduction) use observation frequencies from 111~MHz to 1400~MHz. Therefore, in order to compare the sensitivity when searching for pulsars in different surveys, the sensitivity estimates from the original works must be translated to a frequency of 111 MHz, taking into account the expected spectral index of the pulsar spectrum. All other things being equal, the sensitivity when searching for pulsars in the Galactic plane is 2-3 times lower than outside the plane due to the difference in background temperatures in the Galactic plane and outside it. On the other hand, it is in the plane of the Galaxy that the vast majority of known pulsars are located. Therefore, in some of the surveys listed in the Introduction, a large integral accumulation time was used when observing the Galactic plane, and a small accumulation time outside the plane, and therefore the sensitivity in these surveys is reversed high for the Galactic plane, and lower for observations outside the Galactic plane.

\begin{table*}
	\centering
	\caption{Comparison of surveys currently being conducted to search for pulsars}
	\label{tab:tab2}
	\begin{tabular}{lllllll}
		\hline
Telescope (dimensions) & $\nu$~(MHz)  & $\Delta \nu$~(MHz) & $\Delta t$(c)& $S_{min}$(mJy) & $S_{111}$(mJy)   & links  \\
	\hline
Effelsberg (100m)$^1$	     & 1360 & 240 & 90/1500	& 0.17/0.05 & 28.7/8.4 &  \citeauthor{Barr2013} (\citeyear{Barr2013})  \\ 
Parks (64m)$^1$	             & 1352 & 340 & 270/4300 & 0.61/0.2 & 101.8/33.4 &  \citeauthor{Keith2010} (\citeyear{Keith2010})  \\ 
Green Bank (100m)$^2$	     & 350 & 50   &	140 & 0.6/3.9(1.34) & 7.3/47(16.6)&  \citeauthor{Boyles2013} (\citeyear{Boyles2013}) \\ 
Arecibo (300m)$^3$	         & 327 & 57 & 60 & 0.3/?? & 2.6/?? & \citeauthor{Deneva2013} (\citeyear{Deneva2013}) \\ 
Pushchino (200$\times$400m)     & 111 & 2.5 &180 &      & 7/18 & this paper  \\ 
		\hline
\multicolumn{7}{|p{17cm}|}{Notes to Table 2: According to \citeauthor{Barr2013} (\citeyear{Barr2013}), the sensitivity of the surveys in Effelsberg and Parks is the same, but the sensitivity estimates taken from the original papers differ by almost 4 times. Our estimates of the expected sensitivity when searching for pulsars in Parks practically coincide with the estimates of sensitivity when searching for pulsars on the 100-meter telescope in Effelsberg; 2) The sensitivity estimate of the Green Bank antenna for observations in the Galactic plane is given based on the temperature of 300 K (\citeauthor{Boyles2013}, \citeyear{Boyles2013}). Apparently, this is an extreme case. In parentheses in Table~\ref{tab:tab2}, the sensitivity value is given, which is obtained based on the fact that the temperature in the plane of the Galaxy is 90~K, which is 3 times higher than the temperature outside the plane of the Galaxy; 3) In observations on Arecibo, the plane of the Galaxy was excluded (galactic latitudes $|b| > 5^{\circ}$ were taken). The sensitivity for Arecibo is taken from Figure 2 "Mock") (\citeauthor{Deneva2013} \citeyear{Deneva2013}).}\\
\hline
	\end{tabular}
\end{table*}

Table~\ref{tab:tab2} accumulates information on all large surveys currently being conducted. In the first column, a radio telescope is noted, on which observations are made to search for pulsars. Columns 2 and 3 show the frequency at which the survey is conducted and the total band of observations. Column 4 gives the accumulation time. Different accumulation times for high, medium and low galactic latitudes were used for surveys in Parks and in Effelsberg. The table shows the accumulation time for high ($|b|> 15^o$) and through the sign "/"\, low ($|b| < 3.5^o$) galactic latitudes. Column 5 shows estimates of the expected best sensitivity in the survey outside the plane and through the sign "/"\, in the plane of the Galaxy. The column contains the sensitivity estimates given in the original papers. If the sensitivity assessment was not given in the article, then the sensitivity calculation was made according to the values given in the work or was taken from the corresponding figures. Column 6 shows the results of recalculation of the sensitivity of the surveys at a frequency of 111~ MHz, assuming that the spectral index of all pulsars is equal to two. In the original works on pulsar surveys, sensitivity estimates were given under the assumption of different $S/N_{min}$ and different ratios of pulse duration to period. In order to be able to compare the surveys, the sensitivity estimates in them were first recalculated under the assumption $S/N_{min}=6, W_e=0.1P$, and then the expected minimum flux density was recalculated to a frequency of 111~MHz. Finally, column 7 contains references to pulsar surveys, from which the sensitivities given in column 5 were taken, or the numbers on the basis of which we made sensitivity estimates.

Table~\ref{tab:tab2} shows that the expected sensitivity of LPA is inferior to the sensitivity of the 300-meter Arecibo telescope in observations outside the Galactic plane. It is also inferior to the sensitivity of the 100-meter telescope in Effelsberg and, apparently, the 64-meter telescope in Parks when observed in the plane of the Galaxy ($|b| < 3.5^{\circ}$). In all other cases, the expected sensitivity when searching for pulsars on LPA should be comparable or better than in the other cases considered.

For a test search for pulsars, a declinations of $21^{\circ}23^\prime -42^{\circ}08^\prime$ was taken. A total of 24 days of observations were processed. Every day, approximately 400,000 periodic signals were detected in the data, from which digital filters allow you to select several hundred objects that were analyzed manually. During the search, a direct addition of periods was carried out with a search from 0.5 s to 15 s and a search for DM in the range of 0-200 pc /cm$ ^3$. "Short data" was used for the search. A pulsar was considered found if it was detected for at least 3 days, at least once with $S/N>6$.

According to ATNF (https://www.atnf.csiro.au/research/pulsar), 77 pulsars with a period of more than 500~ms and DM up to 200~pc/cm$^3$ fall into the survey area. The search program "blind"\, the search found 27 known pulsars. Data on them are given in the table~3. Column 1 shows the name of the source in the J2000 agreement. In parentheses is the name in the B1950 agreement, if it is used for this pulsar. Columns 2 and 3 contain period from the ATNF catalog and estimates of pulsar periods obtained by us. In columns 4 and 5, estimates of pulsar flux density at frequencies 102.5~MHz (\citeauthor{Malofeev2000} \citeyear{Malofeev2000}) and 400~MHz (ATNF) are written out. Column ~6 gives a rough estimate of the expected pulsar flux density at a frequency of 111 MHz. The estimate was made based on the flux density at 102.5~MHz, if there were such measurements, and based on the flux density at 400~MHz, if there were no measurements at 102.5~MHz. When recalculating, the spectral index $\alpha = 2$ was assumed. If there is an asterisk after the pulsar name, then there is a comment to this source in the notes after Table~\ref{tab:tab3}. Column 7 shows the half-width of the pulsar pulse taken from ATNF. Column 8 shows how many times the pulsar was detected in 24 processed days.

\begin{table*}
	\centering
	\caption{Known pulsars detected during the "blind" search.}
	\label{tab:tab3}
	\begin{tabular}{llllllll}
		\hline
Name & $P_{ATNF}$(c)   & $P_{survey}$(c)   & $S_{102}$(mJy) & $S_{400}$(mJy)& $S_{111}$(mJy) & $W_{50}$(mc) &  N \\
	\hline
PSR J0048+3412 (B0045+33) & 1.21709         &     1.2171        & 88             &     2.3      &    75          &    21.7      &      17  \\ 
PSR J0323+3944 (B0320+39) & 3.03207         &     3.0326        & 230            &    10.8      &    196         &    42.7      &      23  \\ 
PSR J0528+2200 (B0525+21) & 3.74554         &     3.7443        & 100            &     57       &    85          &   185.5      &      21  \\ 
PSR J0611+3016*           & 1.41209         &     1.4120        &                &     1.4      &    18.9        &              &      22  \\ 
PSR J0613+3731            & 0.61920         &     0.6190        &                &     1.6      &    21          &    11        &      18  \\ 
PSR J0754+3231 (B0751+32) & 1.44235         &     1.4422        & 49             &      8       &    42          &   12.2       &       7  \\ 
PSR J0826+2637 (B0823+26) & 0.53066         &     0.5306        & 620            &     73       &    529         &    5.8       &      18  \\ 
PSR J0943+4109            & 2.22949         &     2.2302        &                &     8.6      &    112         &              &      11  \\ 
PSR J1238+2152            & 1.11859         &     1.1181        & 60             &      2       &     51         &              &      17  \\ 
PSR J1239+2453 (B1237+25) & 1.38245         &     1.3822        & 260            &     110      &    222         &   51.1       &      23  \\ 
PSR J1532+2745 (B1530+27) & 1.12484         &     1.1246        & 94             &      13      &    80          &   25.7       &      19  \\ 
PSR J1741+2758            & 1.36074         &     1.3608        & 30             &       3      &    26          &     7        &       7  \\ 
PSR J1758+3030            & 0.94726         &     0.9472        & 60             &     8.9      &    51          &    27        &      23  \\ 
PSR J1813+4013 (B1811+40) & 0.93109         &     0.9311        &                &      8       &   104          &   12.2       &      22  \\ 
PSR J1821+4145*           & 1.26179         &     1.2620        &                &     2.6      &   35           &              &      20  \\ 
PSR J1907+4002 (B1905+39) & 1.23576         &     1.2355        &                &      23      &   299          &   58.5       &      24  \\ 
PSR J1921+2153 (B1919+21) & 1.33730         &     1.3371        & 1900           &      57      &   1620         &   30.9       &      24  \\ 
PSR J2018+2839* (B2016+28)& 0.55795         &     0.5580        & 260            &     314      &    222         &   14.9       &      24  \\ 
PSR J2055+2209 (B2053+21) & 0.81518         &     0.8152        &                &      9       &   117          &   16.9       &      16  \\ 
PSR J2113+2754 (B2110+27) & 1.20285         &     1.2030        & 130            &     18       &   111          &    13        &      24  \\ 
PSR J2139+2242            & 1.08351         &     1.0835        & 30             &              &    26          &   91         &      22  \\ 
PSR J2157+4017 (B2154+40) & 1.52527         &     1.5250        & 200            &     105      &   171          &   38.6       &      21  \\ 
PSR J2207+40              & 0.63699         &     0.6370        &                &     3.8      &    49          &              &      22  \\ 
PSR J2227+3036            & 0.84241         &     0.8423        &                &     2.4      &    31          &              &      17  \\ 
PSR J2234+2114            & 1.35875         &     1.3587        &  35            &     2.6      &    30          &   43         &      17  \\ 
PSR J2305+3100 (B2303+30) & 1.57589         &     1.5758        &                &     24       &   312          &   17.4       &      23  \\ 
PSR J2317+2149 (B2315+21) & 1.44465         &     1.4445        & 100            &     15       &    86          &   20.2       &      24  \\ 
\hline
\multicolumn{8}{|p{17cm}|}{Notes to Table 3: PSR J0611+3016 was observed in a 430 MHz survey made in Arecibo (\citeauthor{Camilo1996} \citeyear{Camilo1996}). PSR J1821+4145 was observed in a 350~MHz survey done in Green-Bank (\citeauthor{Stovall2014} \citeyear{Stovall2014}). PSR J2018+2839 - the estimation of the flux density at 111 MHz is formal, since the spectrum between the frequencies of 102.5 and 400 MHz is inverted.}\\
\hline
	\end{tabular}
\end{table*}

Table~\ref{tab:tab3} shows that the accuracy of determining the pulsar period, for most of the pulsars found, is better than one unit in the third decimal place. This corresponds to the expected accuracy, determined by the record length of 180 seconds, on which the search was conducted. The real sensitivity obtained during the processing of monitoring data for pulsars, according to Table~\ref{tab:tab3}, is difficult to estimate. Consider column 6 of Table~\ref{tab:tab3}. Weak detected pulsars lying in the extragalactic plane have an expected flux density of 20-30 mJy. Weak pulsars lying in the galactic plane have an expected flux density of 80-100 mJy. Both estimates of the flux density are 3 times greater than the sensitivity limit expected in the survey. Thus, the practical confirmed sensitivity is several times lower than expected.

In addition to the known pulsars, 7 new pulsars missing from the ATNF catalog were discovered. The pulsar was considered open if: a) the $S/N$ was greater than 6, at least on one of the processed days; b) the average profiles with approximately the same height are visible on the recordings with a double period; c) the pulsar is detected in the recordings for at least 3 days out of 24 processed; d) the observed S/N from DM has a pronounced maximum. Table~\ref{tab:tab4} shows the characteristics of these pulsars. The first column contains the name of the source in the J2000 agreement. Columns 2 and 3 give the coordinates of the pulsar in right ascension and declination. The accuracy of the coordinates of pulsars J0146+3104, J0928+3037, J1242+3938, J1721+3524 in the right ascension $\pm 40^s$, in declination $\pm 10^\prime$. Pulsars J0220+3622, J0303+2248, J0421+3240 have the corresponding accuracy of $\pm 60^s$ and $\pm 15^\prime$. Columns 4-6 list the characteristics of the pulsar: period, DM, half-width of the average profile. Column 7 shows how many times the pulsar was found during the 24 processed days.

\begin{table*}
	\centering
	\caption{ Characteristics of new pulsars}
	\label{tab:tab4}
	\begin{tabular}{lllllll}
		\hline
Name & $\alpha_{2000}$ & $\delta_{2000}$ & P(s) & DM(pc/cm$^3$) & $W_{0.5}$(ms) & N \\
	\hline
J0146+3104 &$01^h46^m15^s$  & $31^o04^\prime$ & 0.9381  & 24-26	       &      20	   &    7         \\
J0220+3622 &$02^h20^m50^s$  & $36^o22^\prime$ & 1.0297  & 30-50	       &     220	   &    8         \\ 
J0303+2248 &$03^h03^m00^s$  & $22^o48^\prime$ & 1.207   & 15-25	       &      50	   &    4         \\ 
J0421+3240 &$04^h21^m30^s$  & $32^o40^\prime$ & 0.9005	& 60-90	       &     400	   &    4         \\ 
J0928+3037 &$09^h28^m43^s$  & $30^o37^\prime$ & 2.0919	& 20-24	       &     50	       &   16         \\ 
J1242+3938 &$12^h42^m34^s$  & $39^o38^\prime$ & 1.3100	& 25-27	       &     35	       &   14         \\ 
J1721+3524 &$17^h21^m57^s$  & $35^o24^\prime$ & 0.8219	& 19-25	       &     60	       &   18         \\
		\hline
	\end{tabular}
\end{table*}

The data in the Table~\ref{tab:tab4} is preliminary. Currently, observations are being carried out to clarify the parameters (DM and period) of the detected objects on an installation made specifically for the study of pulsars. Pulsars J0146+3104, J0220+3622, J0421+3240, J1242+3938, J1721+3524 have already been confirmed by observations on LPA1. For them, the clarification of the period is no worse than up to the seventh decimal place (\citeauthor{Malofeev2015} \citeyear{Malofeev2015}).

We pay special attention to pulsars J0220+3622, J0421+3240. These pulsars have broad average profiles that can take up about half of the period.

For the new pulsars, the Fig.1 show dynamic spectra, average profiles, and $S/N$ from $DM$ for the day of observations. "Long data" was used to obtain the drawings. The drawings of the dynamic spectrum and the average profile are made with a double period. Dynamic spectra were constructed using "long data". The pulsars found are weak, so averaging was carried out within the dynamic spectrum by frequencies and/or time (see the comments to the figure). The minimum frequencies within the observation band correspond to the upper part of the dynamic spectrum. The maximum frequencies correspond to the lower part of the spectrum.

\begin{figure*}
\begin{center}
		\includegraphics[width=17cm]{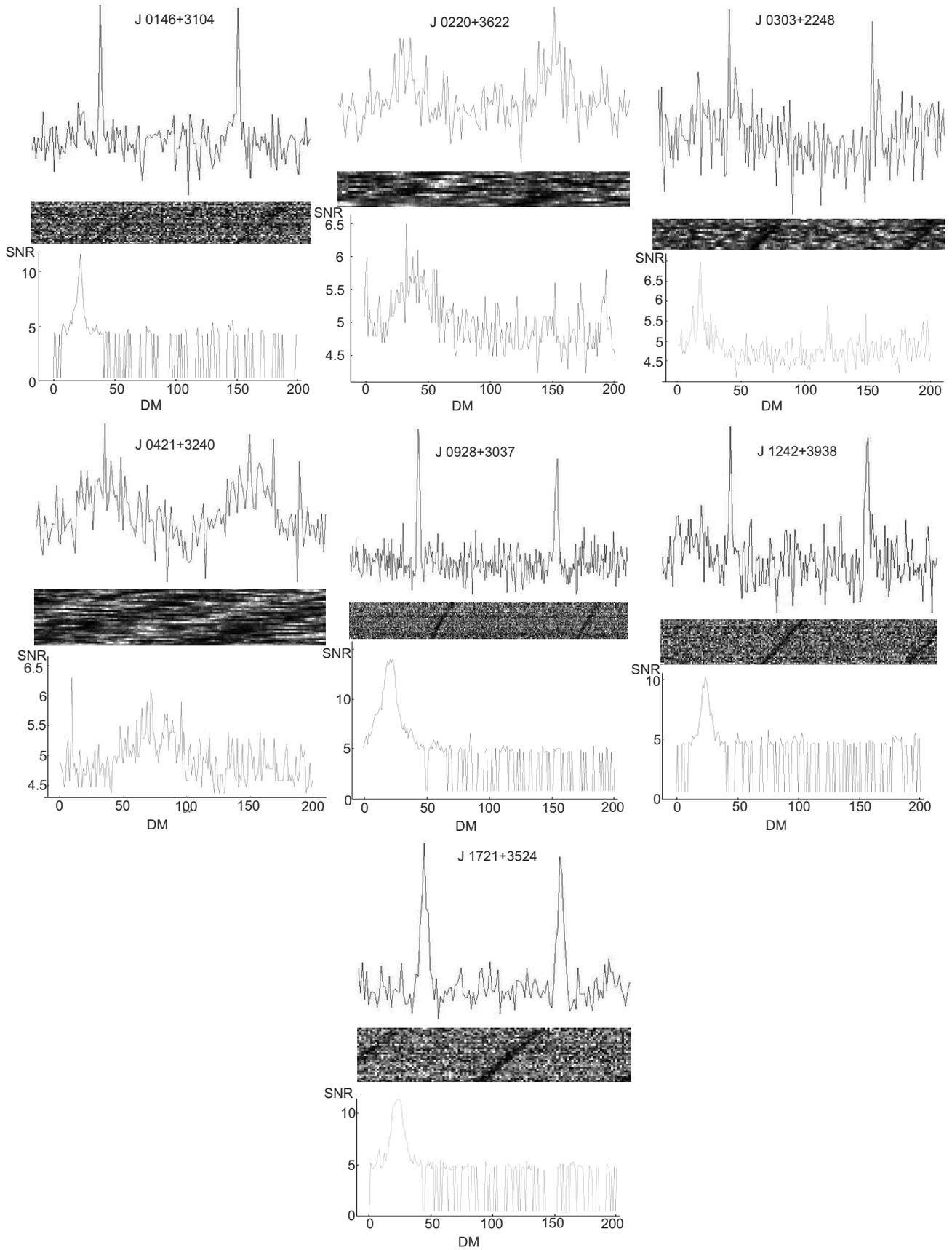}
    \caption{Average profiles, dynamic spectra and (S/N) from DM of new pulsars}
\end{center}
\end{figure*}

\section{DISCUSSION OF THE RESULTS AND THE CONCLUSION}

Table~\ref{tab:tab1} shows a theoretical estimate of the expected sensitivity of the survey when searching for pulsars on LPA LPI. Processing a real survey shows that the sensitivity is noticeably worse than the calculated one. There are a number of factors, taking into account which will allow us to approach the calculated sensitivity.

Firstly, the maximum sensitivity when searching for pulsars will be achieved when the pulse duration of the pulsar is equal to the sampling. If the sampling in the raw data is less than the expected pulse duration of the pulsar, then an additional averaging can be performed within the expected pulsar period and the maximum S/N can be obtained. Table~\ref{tab:tab3} shows that approximately 80\% of all detected known pulsars have a half-width of the average pulsar pulse profile less than 30~ms, therefore, taking into account our sampling of 100~ms, the loss in the S/N when processing these pulsars was 1.5 times or more. First of all, S/N losses will affect the detection of the weakest pulsars. These losses can be avoided by processing "long data". A sampling of 12.5~ms is sufficient to obtain the maximum S/N when searching for the vast majority of known second pulsars, and, most likely, it will be sufficient when searching for new pulsars.

Secondly, in the meter wavelength range, scintillation of radio sources is observed on the interplanetary plasma and on the ionosphere of the Earth. According to earlier studies conducted by (\citeauthor{Artyukh1982} \citeyear{Artyukh1982}), the median value of the confusion effect of scintillating radio sources at 102 MHz is 0.14~Jy, which is close to the fluctuation sensitivity of LPA2. Scintillation of compact radio sources is constantly observed in real primary recordings. Early studies at the LPA LPI showed that the density of detected compact (scintillating) radio sources in the sky is about 1 source per 1 sq.deg. (\citeauthor{Artyukh1996} \citeyear{Artyukh1996}), which is comparable to the size of the radiation pattern of LPA. Since scintillation is a random process, they increase the width of the noise track and thereby reduce the sensitivity of the survey when searching for pulsars. Note also that the characteristic scintillation time in the meter wavelength range is about 0.5s. Therefore, there is a problem of subtracting the background signal from the recording. The scintillation detection area is wide. Observations under the "Space Weather" program\, on the LPA show that the zone of increased scintillating, with the current sensitivity of the LPA2 antenna, can extend for 12-18 hours, depending on the time of year  (\citeauthor{Chashei2015} \citeyear{Chashei2015}). Ionospheric scintillating, excluding magnetic storms, takes about one hour in the morning and in the evening. The characteristic time of ionospheric scintillation starts from a few seconds (\citeauthor{Chashei2006} \citeyear{Chashei2006}), which, as for interplanetary scintillation, can lead to problems subtracting the background from the recording.

Therefore, when searching for pulsars to achieve guaranteed maximum sensitivity, it makes sense to choose only 5-6 night hours in the recordings.

Thirdly, during the search, about a third of the known pulsars ($P>0.5$~c) in investigated area were detected. About another third of the known pulsars not detected during the search are most likely too weak to detect them. The expected flux density of these pulsars, extrapolated from ATNF estimates of flux densities, may be less than 20-30 mJy at a frequency of 111~MHz. These are pulsars J0540+3207, J0546+2441, J0555+3948, J0947+2742*, J1503+2111, J1720+2150, J1746+2245, J1746+2540, J1900+3053, J1903+2225, J1912+2525, J1913+3732, J1929+2121, J1931+3035, J1937+2950, J1939+2449, J1946+2224, J1949+2306, J1953+2732, J2007+3120, J2010+2845, J2015+2524, J2036+2835, J2151+2315*, J2155+2813*. The pulsars listed above, with the exception of those marked with the sign '*', are located in the plane of the Galaxy, where the sensitivity of the survey to search for pulsars is the worst. Some pulsars are X-ray or RRATs objects. These are pulsars J1308+2127, J1605+3249, J2225+3530. Nevertheless, there remain approximately 20-25 known pulsars, which by all indications should have been detected during the search, but were not detected. In particular, some of these pulsars were previously observed on LPA at a frequency of 102.5 MHz. These are pulsars J0417+3545 ($S_{102}=46$~mJy), J0927+2347* ($S_{102}=30$~mJy), J0943+2256* ($S_{102}=77$~mJy), J1649+2533 ($S_{102}=60$~mJy), J1652+2651 ($S_{102}=40$~mJy), J1920+2650 ($S_{102}=30$~mJy), J1948+3540 ($S_{102}=60$~mJy), J2002+3217 ($S_{102}=80$~mJy), J2002+4050 ($S_{102}=170$~mJy), J2008+2513 ($S_{102}=60$~mJy), J2030+2228 ($S_{102}=22$~mJy), J2037+3621 ($S_{102}=34$~mJy), J2212+2933* ($S_{102}=50$~mJy), J2307+2225* ($S_{102}=30$~mJy) \cite{Malofeev2000}. Pulsars marked with the sign '*' are located at great distances from the plane of the Galaxy. Approximately half of the pulsars listed in this paragraph were found either at $S/N<6$, or less than three times, or at multiple harmonics, and therefore, when searching for new pulsars, it would be eliminated. Currently, methodological work is being carried out to improve the search program.

There are a number of other factors that it makes sense to mention in the work: 1) at low declinations, there is a large amount of interference sitting in the south direction and being industrial interference; 2) in spring and early summer, a lot of records disappear due to thunderstorms; 3) the 111~MHz range on which observations are carried out is not protected on primary and secondary bases, so it is regularly used interference appears to be associated with mobile services. All these factors lead to the fact that approximately 20-25\% of all records cannot be processed.

Concluding the paper, we note that the main advantages of LPA is its high effective area, and therefore high sensitivity, the possibility of simultaneous observations in many beams, the possibility of daily monitoring. In other current pulsar search surveys (see Table~\ref{tab:tab2}) high sensitivity of observations is achieved due to a wide frequency band, and due to the fact that the temperature of the system ($T_{sys}$) decreases with increasing frequency of observations. The possibility of daily monitoring of the entire sky in these surveys is excluded.

The search for pulsars in the daily monitoring data on LPA LPI is especially advantageous for detecting rare objects: flashing pulsars, in which long periods of relative rest are replaced by a significant increase in the observed flux density, pulsars of the RRATs type, pulsars of the Geminga type, pulsars with nullings, pulsars with giant pulses. Separately, we note close pulsars, which, due to interstellar scintillation, can significantly change the observed flux density from day to day, as well as pulsars with very steep spectra ($\alpha > 2.5$).

In conclusion, we note that when processing half of the available declination area and using "short data", 7 new pulsars were detected. Taking into account the apparent loss of sensitivity in the search, the nature of which is not entirely clear, and taking into account the fact that "long data" will eventually be processed, we can expect the detection of at least several dozen new pulsars in the monitoring data of LPA LPI.

\section*{Acknowledgments}

We express our gratitude to V.M. Malofeev for the preliminary reading of the manuscript and a number of comments that made it possible to improve its text, as well as L.B. Potapova and G.E. Tyulbasheva for their help in the design of the article and pics. This work was supported by the program of the Division of Physical Sciences of the Russian Academy of Sciences “Transient and explosive processes in astrophysics.”

\bsp	
\label{lastpage}

\begin{thebibliography}{99}
\bibliographystyle{unsrt} 

\bibitem [Hewish(1968)]{Hewish1968} Hewish A., Bell S.~J., Pilkington J.~D.~H., Scott P.~F., Collins R.~A., 1968, Natur, 217, 709. doi:10.1038/217709a0

\bibitem[Large(1971)]{Large1971} Large M.L., Vaughan A.E., 1971, MNRAS, bf 151, 277

\bibitem[Manchester(1978)]{Manchester1978} Manchester R.N., Lyne A.G., Taylor J.H., Durdin J.M., Large M.I., Little A.G., 1978, MNRAS, 185, 409

\bibitem[Davies(1972)]{Davies1972} Davies J.G., Lyne A.G., Seiradakis J.H., 1972, Nature, 240, 229

\bibitem[Davies(1973)]{Davies1973} Davies J.G., Lyne A.G., Seiradakis J.H., 1973, Nature Phys. Sci., 244, 84

\bibitem[Damashek(1978)]{Damashek1978} Damashek M., Taylor J.H., Hulse R.A., 1978, Astroph. J., 225, L31

\bibitem[Deneva(2013)]{Deneva2013} Deneva J. S., Stovall K., McLaughlin M.A., Bates S.D., Freire P.C.C., Martinez J.G., Jenet F., Bagchi M., 2013, Astroph. J., 775, 51

\bibitem[Boyles(2013)]{Boyles2013} Boyles J., Lynch R.S., Ransom S.M., et.al., 2013, Astroph. J., 763, 80

\bibitem[Barr(2013)]{Barr2013} Barr E.D., Champion D.J., Kramer M., Eatough R.P., et.al., 2013, MNRAS, 435, 2234

\bibitem[Keith(2010)]{Keith2010} Keith M.J., Jameson A., van Straten W., Bailes M., et.al., 2010, MNRAS, 409, 619

\bibitem[Coenen(2013)]{Coenen2013} Coenen T., 2013,  PhD thesis

\bibitem[Malofeev(2000)]{Malofeev2000} Malofeev V.M., Malov O.I., Shchogoleva N.V., 2000, 
A.Zh., 77, 499

\bibitem[Malofeev(2015)]{Malofeev2015} Malofeev V.M. et al., Astron. Astroph. (in print)

\bibitem[Rickett(1977)]{Rickett1977} Rickett B.J., 1977, Annual Review of Astron.Astroph., 15, 479

\bibitem[Vitkevich(1969)]{Vitkevich1969} Vitkevich V.V., Alekseev Y.A., Zhuravlev V.F., Shitov Yu.P., 1969, Nature, 224, 49

\bibitem[Tartle(1962)]{Tartle1962} Turtle A.J., Baldwin J.E., 1962, MNRAS, 124, 459

\bibitem[Tyulbashev(2015a)]{Tyulbashev2015a} Tyul'bashev S.A., Tyul'bashev S.A., ATsir

\bibitem[Tyulbashev(2015b)]{Tyulbashev2015b} Tyul'bashev S.A., Tyul'bashev S.A., ATsir

\bibitem[TyulbashevV(2015)]{TyulbashevV2015} https://github.com/vtyulb/BSA-analytics

\bibitem[Camilo(1996)]{Camilo1996} Camilo F., Nice D.J., Shrauner J.A., Taylor J.H., 1996, Astroph.J., 469, 819

\bibitem[Stovall(2014)]{Stovall2014} Stovall K., Lynch R.S., Ransom S.M., Archibald A.M., et.al., 2014, Astroph. J., 791, 67

\bibitem[Artyukh(1982)]{Artyukh1982} Artyukh V.S., Shishov V.I., 1982, A.Zh, 59, 896

\bibitem[Artyukh(1996)]{Artyukh1996} Artyukh V.S., Tyul'bashev S.A., 1996, A.Zh., 73, 661

\bibitem[Chashei(2015)]{Chashei2015} Chashei I.V., Shishov V.I., Tyul'bashev S.A., Subaev I.A., Oreshko V.V., Logvinenko S.V., 2015, Solar Physics, 

\bibitem[Chashei(2006)]{Chashei2006} Chashei I.V., Shishov V.I., Vlasov V.I., Tyul'bashev S.A., Subaev I.A., Shutenkov V.R, 2006, IzSSR, 70, 1542

\end{thebibliography}
\end{document}